\documentclass[12pt]{iopart}
\usepackage[utf8x]{inputenc}
\usepackage{graphicx}
\usepackage{hyperref}
\usepackage{cleveref}
\usepackage{bm}
\usepackage{color}

\usepackage{iopams}  
\begin{document}

\title{Nuclear matter at finite temperature and static properties of proto-neutron star}

\author{Debashree Sen}

\address{Department of Physical Sciences \\Indian Institute of Science Education and Research Berhampur,\\ Transit Campus, Government ITI, 760010 Berhampur, Odisha, India}
\ead{debashree@iiserbpr.ac.in}
\vspace{10pt}
\date{\today}

\begin{abstract}

With the effective chiral model, the finite temperature properties of nuclear matter  have been studied at different temperatures. For symmetric nuclear matter, I particularly focused on the possibility of liquid-gas phase transition at low temperature and density. The critical temperature obtained in this context, is consistent with the experimental and empirical findings. The free energy and entropy variation are also studied for different values of temperature. A few asymmetric nuclear matter properties like the equation of state and the speed of sound with respect to temperature are also examined. The work is also extended to obtain the equation of state $\beta$ stable nuclear matter at finite temperature. For the neutrino free case, the various static proto-neutron star properties are computed for a wide range of temperature, relevant to proto-neutron stars. For all the values of temperature, the obtained estimates of maximum gravitational mass are found to be in good agreement with the observational constraints specified from massive pulsars like PSR J0348+0432 and PSR J0740+6620. The results of surface redshift for all the temperature also satisfy the maximum surface redshift constraints from EXO 07482-676, 1E 1207.4-5209 and RX J0720.4-3125.

\end{abstract}

%
%
%
%

\section{Introduction}
\label{intro}

 The birth of a proto-neutron star (PNS) is an outcome of the gravitational collapse of the core of a very massive ($\geq 8 M_{\odot}$) and highly-evolved star. The collapse process is known as a supernova explosion (type-II) \cite{Glendenning,Lattimer}. Over short timescale a PNS undergoes several stages and cools down by the emission of neutrinos to form a stable neutron star (NS) of negligible temperature at MeV scale ($T=0$). The various stages of evolution of a PNS are studied in great details in many works like \cite{All}. A newly born PNS is composed of neutrino-trapped, very dense $\beta$ equilibrated matter at temperature as high as 50 MeV \cite{Dex}. The neutrino-trapped stage is short-lived (about 10 to 20 seconds) as the star eventually cools down fast to reach a quasi-equilibrium state by the emission of neutrinos \cite{All,Dex,Yu}. 
 
 The structural properties of PNSs depend largely on the equation of state (EoS) and composition of nuclear matter at high density ($(5-10) \rho_0$ ; where, $\rho_0=0.16$fm$^{-3}$ is the normal nuclear matter density) and finite temperature (upto 50 MeV). Since the composition of such matter is largely determined by the characteristics of strong interactions which is at present largely unknown, there is a lot of uncertainty pertaining to the EoS of nuclear matter at this density and temperature regime. Matter at such conditions is characterized by isospin asymmetry and plays important role not only to describe the composition of NS cores but also to explain heavy-ion collision dynamics, nuclear structure, core-collapse supernova explosion and binary compact star mergers \cite{Zhang,Burgio,Oertel,Bharat,Zhang2016}. At present the study of dense asymmetric nuclear matter (ANM) at finite temperature is an active area of research at various facilities like HIE-ISOLDE at CERN, SPIRAL-2 at GANIL, TRIUMF in Canada, RIKEN in Japan, FRIB at MSU, RHIC at Brookhaven National Laboratory and the LHC at CERN.
 
 It is therefore that one rely on theoretical modeling of proto-neutron star matter (PNSM) to understand its composition and compute the EoS on its basis. A lot of work has been done in this regard to comprehend successfully the temperature dependence of nuclear matter properties \cite{All,Dex,Yu,Zhang,Burgio,Oertel,Bharat,Zhang2016,Khvorostukhin,Zhou,Tan,Baldo1999,Burgio2010,
Wang2000,Malheiro,Hong2016,Nicotra,Sahu2004,Du,Camelio,Jena} using a wide variety of approaches. Based on microscopic treatments, two and three body nuclear potentials were constructed using chiral effective theory \cite{Carbone2018,Wellenhofer2015} to compute the nuclear matter properties at finite temperature while nuclear dipolar strength was successfully calculated by adopting finite-temperature relativistic time blocking approximation \cite{Wibowo}. Self-consistent Hartree-Fock approach has been utilized in \cite{Sauer,BALi,Carbone2011,Lourenco}, considering different interactions to investigate finite temperature properties of nuclear matter. Brueckner-Hartree-Fock approach has also been extensively adopted for the same purpose \cite{Lu,Burgio2010,Nicotra,Shang2020,Figura,Wei2020,Fortin2018,Baldo2016} while liquid drop type of model was earlier introduced in \cite{LatSwes}. Phenomenological models based on nuclear condensation were adopted in \cite{Jaqaman,Kapusta} while relativistic mean field (RMF) models were also successfully used for the same purpose \cite{Shen1998,Shen2011}. Beyond mean field treatments also helped to understand nuclear matter properties at finite temperature \cite{Friedman,Baldo1999,Rios2008,Soma}. Ref. \cite{LatPrac2016} provides a good review of the various approaches adopted in this regard. In the present work I employ the well-tested effective chiral model \cite{Sahu2004,TKJ} to study the finite properties of nuclear matter at different density regimes. This phenomenological model is based on chiral symmetry and RMF treatment, The masses of the nucleons and the scalar and vector mesons are generated dynamically. The model has a very few free parameters and they are fixed by satisfying the saturated nuclear matter (SNM) properties \cite{Sahu2004,TKJ,TKJ2,TKJ3}. The same model was employed by \cite{Sahu2004} to understand a few finite temperature properties of matter at low density. In several works \cite{Baldo1999,Nicotra,BurgioNew} that also investigated the finite temperature properties of nuclear matter, the entire interaction involves temperature dependence. However, similar to \cite{All,Yu,Zhang,Hong2016,Sahu1993,Wang2000,Bharat}, in \cite{Sahu2004} the temperature effects were invoked through the kinetic part of the interaction which also successfully explained the finite temperature properties of nuclear matter at low density. In the present work I consider same approach with the same model as \cite{Sahu2004} to examine the thermal properties of nuclear matter more rigorously at different density domains. It is well-known that nuclear matter properties are best determined at low density mainly around $\rho_0$. The phenomenon of liquid-gas phase transition is of great interest in this domain of study \cite{Sahu2004,Kupper,Friedman,
Jaqaman,Bandyopadyay,Song,Song2,Muller,Panda,Bharat,Zhang2016,Baldo1999,Wang2000,
Malheiro,Jena}. The isotherms at low density manifest a Van der Waal like interaction indicating the coexistence of both liquid and gaseous phases upto a certain density depending on the temperature. At very low temperature, a typical unphysical region in each isotherm is observed over a density range. This region is indicated by the negative values of binding energy per particle and pressure. With increasing temperature, the density range of this region decreases and finally at a particular temperature the values of binding energy per particle and pressure are no longer negative indicating the complete transition of liquid phase to a gaseous one. This transition temperature is called the critical temperature $T_c$. A lot of experimental endeavors have determined the range of $T_c$ to be within (10-20) MeV \cite{Sharma,Kaijpper,Yang,Elliott,Takatsuka,Holt,Karnaukhov,Bharat,Sahu2004}. It is also known that liquid-gas phase transition can also be attained with ANM for low values of the asymmetry parameter $\alpha$ and temperature \cite{Wang2000,Sahu2004}. 

 With the increase in density, the matter becomes isospin asymmetric and properties of such matter differs from that of SNM. In this work I have therefore treated matter as asymmetric and re-computed some of the properties of such matter like the EoS and the speed of sound. As mentioned earlier that nuclear matter at high density is highly applicable to account for the composition of PNSs and consequently their structural properties \cite{All,Pons,BurgioNew,Lenka,Roark,Bharat,Lu,Strobel,Dex,Tan,Burgio2010, Burgio2011,Chen,Hong2016,Jena,Yu,Nicotra}. In the present work I considered $\beta$ equilibrated matter at finite temperature and for the neutrino untrapped scenario \cite{All,Chen,Bharat,Tan}, I investigated the role of temperature to determine the static properties of PNSs like gravitational and baryonic mass, radius and the surface redshift. I have also compared the obtained results with the observational constraints like the maximum mass estimates from massive pulsars like PSR J0348+0432 ($M = 2.01 \pm 0.04~ M_{\odot}$) \cite{Ant} and PSR J0740+6620 ($2.14^{+0.10}_{-0.09}~ M_{\odot}$ (68.3\%) and $2.14^{+0.20}_{-0.18}~ M_{\odot}$ (95.4\%)) \cite{Cromartie} and the maximum surface redshift obtained from EXO 07482-676 ($Z_s=0.35$) \cite{Cottam2002}, 1E 1207.4-5209 ($Z_s=(0.12-0.23)$) \cite{Sanwal} and RX J0720.4-3125 ($Z_s=0.205^{+0.003}_{-0.006}$) \cite{Hambaryan}.
 
 The primary aim of this work is to test the validity of the effective chiral model \cite{Sahu2004,TKJ} in the different density regimes at finite temperature in order to understand the properties of isospin symmetric matter at sub-nuclear density as well as $\beta$ stable matter at large densities relevant to PNSs. At low density, I have discussed liquid-gas phase transition and calculated the properties like the free energy, entropy and the speed of sound in this connection along with the EoS. I also extended the work further with the same model to understand the properties of $\beta$ stable matter at large density and finite temperature relevant to PNSs. The structural properties of PNSs are computed in static conditions and compared with the recent constraints obtained from certain pulsars.
 
 This paper is planned as follows. In the next section \ref{Model}, I describe the aspects of the model considered and the EoS at both zero (section \ref{EOS_0}) and finite (section \ref{EOS_T}) temperature. I also discuss the mechanism to treat SNM, ANM and $\beta$ equilibrated matter separately and present the parameter set of the model adopted and its connection to the SNM properties at $T=0$. I also discuss the methodology to compute static PNS properties in the same section \ref{Structure}. I present the obtained results and analysis in the next section \ref{Results} and finally conclude in the closing section \ref{conclusion} of the paper.

\section{Formalism}
\label{Formalism}

\subsection{Effective Chiral Model}
\label{Model}

 The effective Lagrangian density \cite{Sahu2004,TKJ} for the effective chiral model is given by
\begin{eqnarray}
\hspace*{-0.5in}\mathcal{L} &=& \overline{\psi} \Biggl[ \left(i \gamma_{\mu} \partial^{\mu} - g_{\omega}~ \gamma_{\mu} \omega^{\mu} -\frac{1}{2} g_{\rho}~ \overrightarrow{\rho_{\mu}} \cdot \overrightarrow{\tau} \gamma^{\mu} \right)-g_{\sigma} \left(\sigma + i \gamma_5 \overrightarrow{\tau} \cdot \overrightarrow{\pi} \right) \Biggr] \psi \nonumber \\
\hspace*{-0.5in}&+& \frac{1}{2} \left(\partial_{\mu} \overrightarrow{\pi} \cdot \partial^{\mu} \overrightarrow{\pi} + \partial_{\mu} \sigma ~ \partial^{\mu} \sigma \right) -{\frac{\lambda}{4}} \left(x^2-x_0^2\right)^2 - \frac{\lambda B}{6} (x^2-x_0^2)^3 - \frac{\lambda C}{8}(x^2-x_0^2)^4 \nonumber \\
\hspace*{-0.5in}&-& \frac{1}{4}F_{\mu\nu}F^{\mu\nu} 
+\frac{1}{2} {g_{\omega}}^2~x^2~\omega_\mu \omega^\mu 
- \frac{1}{4}~\overrightarrow{R_{\mu\nu}} \cdot 
\overrightarrow{R^{\mu\nu}}+\frac{1}{2}~m_\rho^2 ~\overrightarrow{\rho_\mu} \cdot \overrightarrow{\rho^\mu}  
\protect\label{Lagrangian}
\end{eqnarray}

where, $\psi$ is the nucleon isospin doublet. In the mean field (MF) treatment, $\left\langle\pi\right\rangle=0$ and the mass of the pions is $m_{\pi}=0$. Therefore  only the non-pion matter \cite{Sahu2004,TKJ,Sen,Sen2,Sen3,Sen4} is taken into account. The interaction between the nucleons (N=n,p) takes place via the scalar $\sigma$, the vector $\omega$ (783 MeV) and the isovector $\rho$ (770 MeV) mesons, having respective coupling strengths as $g_{\sigma}, g_{\omega}, g_{\rho}$. $B$ and $C$ are the couplings associated with the higher order terms of the scalar field. Their values are determined along with that of $g_{\sigma}$, $g_{\omega}$ and $g_{\rho}$ at saturation density $\rho_0$ \cite{TKJ}. 

 The model is based on spontaneous breaking of the chiral symmetry at ground state as a result of which the scalar field $\sigma$ attains a vacuum expectation value (VEV) $x_0$. Both the scalar and vector fields are chiral invariant as $x^2 = ({\pi}^2+\sigma^2)$ \cite{Sahu2004,TKJ,Sen,Sen2,Sen3,Sen4}. The interaction of the scalar $\sigma$ and the pseudoscalar $\pi$ mesons with the isoscalar vector boson $\omega$ leads to the dynamical generation of the mass term of the $\omega$ field given as $\frac{1}{2} {g_{\omega}}^2~x^2~\omega_\mu \omega^\mu$. This indicates the explicit dependence of the nucleon effective mass on both the scalar and the vector fields \cite{Sahu2004,TKJ,Sen,Sen2,Sen3,Sen4}. Due to symmetry breaking, the mass of the nucleons ($m$) is also dynamically generated along withthat of the scalar and vector mesons \cite{Sahu2004,TKJ,Sen,Sen2,Sen3,Sen4}.
 
 The isospin triplet $\rho$ meson is included to account for the ANM at higher density \cite{Sahu2004,TKJ,Sen,Sen2,Sen3,Sen4,Sahu1993,Sahu2000}. Its coupling strength ($g_{\rho}$) is obtained by fixing the symmetry energy coefficient $J = 32$ MeV at $\rho_0$, given by \cite{Sahu2004,TKJ}

\begin{eqnarray}
J = \frac{C_{\rho}~ k_{F}^3}{12\pi^2} + \frac{k_{F}^2}{6\sqrt{(k_{F}^2 + m^{\star 2})}}
\end{eqnarray}

 where, $C_{\rho} \equiv g^2_{\rho}/m^2_{\rho}$, $m^*$ is the nucleon effective mass and $k_{F}=(6\pi^2 \rho/{\gamma})^{1/3}$ is the Fermi nucleon momenta in terms of total baryon density $\rho$. In addition, the asymmetric parameter also ensures the invocation of neutron-proton asymmetry in the matter. It is defined as \cite{Bharat,Wang2000,Sahu2004}
 
\begin{eqnarray}
\alpha=\frac{\rho_n-\rho_p}{\rho_n+\rho_p}
\label{alpha}
\end{eqnarray}

 The total baryon density ($\rho$) is the sum of individual nucleon densities i.e., $\rho=\rho_n + \rho_p$. The spin degeneracy factor ($\gamma$) and asymmetric parameter ($\alpha$) are (4,0) for SNM and (2,1) for pure neutron matter (PNM).

 At higher density, matter becomes more and more asymmetric as the number of neutrons become exceedingly larger than that of protons. At densities relevant to neutron star matter (NSM) ($\rho \geq 10^{14}$ gm cm$^{-3}$), electron capture and inverse $\beta$ decay are the two important processes that come into play and the matter can be treated as $\beta$ equilibrated matter \cite{All,Glendenning}. At densities ($\lesssim 2 \rho_0$) the Fermi momentum of electrons may equal and even surpass the mass of muons ($m_{\mu}=105.6$~MeV). It is at that time the muons start appearing in NSM. The resultant matter should be in chemical equilibrium in order to attain minimum energy configuration of matter. The nucleon chemical potential is given as \cite{Sahu2004,TKJ} 

 \begin{eqnarray}
\mu_B =\sqrt{k^2+m^{\star 2}}-g_{\omega}{\omega}_0 + g_{\rho}I_{3B}\rho_{03}
\end{eqnarray}

where, suffix B=n,p and $I_{3n}=-1/2$ for neutron and $I_{3p}=+1/2$ for proton. Here `3' denotes for the third component in isospin space because in the MF treatment only the third component survives in case of the isospin triplet $\boldsymbol{\rho}=\rho^{{\pm,0}={1,2,3}}$.

 Also, the NSM should be charge neutral. Therefore the charge neutrality condition should also be imposed as
 
\begin{eqnarray}
\rho_p=\rho_e+\rho_{\mu} 
\end{eqnarray}

 I consider the neutrino free $\beta$ equilibrated matter consisting of the nucleons, electrons and muons as the composition of PNSM for the present work following \cite{All,Chen,Bharat,Tan}. 
 
\subsection{Equation of State at T = 0} 
\label{EOS_0}

 Considering the matter as Fermi gas, at zero temperature  the scalar density is obtained as \cite{Sahu2004,TKJ,Sahu2000}

\begin{eqnarray}
\rho_S=\left\langle\overline{\psi}\psi\right\rangle=\frac{\gamma}{2 \pi^2} \int^{k_F}_0 dk ~k^2 \frac{m^*}{\sqrt{k^2 + {m^{*}}^2}}
\end{eqnarray} 

while the baryon density as

\begin{eqnarray} 
\rho=\left\langle\psi^\dagger\psi\right\rangle=\frac{\gamma}{2\pi^2} \int^{k_F}_0 dk ~k^2  
\end{eqnarray}

The equation of motion of the nucleons and the mesonic fields are calculated in a RMF approach in terms of their mean field values \cite{Sahu2004,TKJ}. They are utilized to obtain the EoS for ANM given as \cite{Sahu2004,TKJ}

\begin{eqnarray} 
\hspace*{-2cm}\varepsilon = \frac{m^2}{8~C_{\sigma}}(1-Y^2)^2 - \frac{m^2 B}{12~C_{\omega}C_{\sigma}}(1-Y^2)^3
+\frac{C m^2}{16 ~C_{\omega}^2~ C_{\sigma}}(1-Y^2)^4 \nonumber \\ \hspace*{-2cm} + \frac{1}{2Y^2}C_{\omega} {\rho^2} + \frac{\gamma}{\pi^2} \int_{0}^{k_F} k^2 \sqrt{(k^2+{m^*}^2)} ~dk + \frac{\gamma}{2\pi^2} \sum_{\lambda= e,\mu^-} \int_{0}^{k_\lambda} k^2 \sqrt{(k^2+{m_\lambda}^2)}~ dk
\protect\label{EoS1cn}
\end{eqnarray}

\begin{eqnarray}         
\hspace*{-2cm}P =-\frac{m^2}{8~C_{\sigma}}(1-Y^2)^2+\frac{m^2 B}{12~C_{\omega}~C_{\sigma}}(1-Y^2)^3
-\frac{C~ m^2}{16~ C_{\omega}^2 C_{\sigma}}(1-Y^2)^4 \nonumber \\
\hspace*{-2cm}+\frac{1}{2Y^2}C_{\omega} {\rho^2} + \frac{\gamma}{3\pi^2} \int_{0}^{k_F} \frac{k^4}{\sqrt{(k^2+{m^*}^2)}}~ dk + \frac{\gamma}{6\pi^2} \sum_{\lambda= e,\mu^-} \int_{0}^{k_\lambda} \frac{k^4}{ \sqrt{(k^2+{m_\lambda}^2)}}~ dk
\protect\label{EoS2cn} 
\end{eqnarray}

 where, $C_i={g_i}^2/{m_i}^2$ are the scaled couplings where $i=\sigma,\omega~\&~\rho$, $m_i$ being the mass of the mesons. It is to be noted that the $\rho$ mesons have no role to play in case of SNM. Also for SNM the last term does not exist. The contribution of the last term due to leptons ($\lambda= e,\mu^-$) is considerable only when $\beta$ equilibrated matter is considered at high density for PNSM.

\subsection{Equation of State at finite temperature} 
\label{EOS_T}

  At finite temperature, the momentum $k$ is a function of temperature \cite{Yu,Zhang,Sahu2004,All,Sahu1993,Jena}. This makes both the scalar and baryonic densities functions of temperature. 
  
  Therefore the scalar density can be written as \cite{Yu,Zhang,Sahu2004,Wang2000,All,Jena,Sahu1993}
 
\begin{eqnarray}
\rho_S=\frac{\gamma}{2 \pi^2} \int^{\infty}_0 dk ~k^2 \frac{m^*}{\sqrt{k^2 + {m^{*}}^2}} (f(T)+\bar{f}(T))
\end{eqnarray} 

 while the baryon density as \cite{Yu,Zhang,Sahu2004,Sahu1993,Jena}

\begin{eqnarray} 
\rho=\frac{\gamma}{2\pi^2} \int^{\infty}_0 dk ~k^2 (f(T)-\bar{f}(T))  
\end{eqnarray}

 where, the nucleon and anti-nucleon distribution functions $f(T)$ and $\bar{f}(T)$ are respectively,
expressed as\cite{Yu,Zhang,Sahu2004,Sahu1993,Jena}

\begin{eqnarray}
f(T) = \frac{1}{exp[(E^* - \nu)/T] + 1}
\end{eqnarray}

and 

\begin{eqnarray}
\bar{f}(T) = \frac{1}{exp[(E^* + \nu)/T] + 1}
\end{eqnarray}

where, $E^* = \sqrt{k^2 + {m^*}^2}$ and the chemical potential is defined as

\begin{eqnarray}
\nu = \mu - g_{\omega}\omega_0 + I_{3B} g_\rho \rho_{03}
\end{eqnarray}

where, the last term does not exist for SNM.

 The EoS at finite temperature for ANM is thus given as \cite{Sahu2004,TKJ}
 
\begin{eqnarray} 
\hspace*{-2cm}\varepsilon = \frac{m^2}{8~C_{\sigma}}(1-Y^2)^2 - \frac{m^2 B}{12~C_{\omega}C_{\sigma}}(1-Y^2)^3
+\frac{C m^2}{16 ~C_{\omega}^2~ C_{\sigma}}(1-Y^2)^4 \nonumber \\ \hspace*{-2cm} + \frac{1}{2Y^2}C_{\omega} {\rho^2} + \frac{\gamma}{\pi^2} \int_{0}^{\infty} k^2 \sqrt{(k^2+{m^*}^2)} (f(T)-\bar{f}(T))~dk \nonumber \\+ \frac{\gamma}{2\pi^2} \sum_{\lambda= e,\mu^-} \int_{0}^{\infty} k^2 \sqrt{(k^2+{m_\lambda}^2)} (f_{\lambda}(T)+\bar{f_{\lambda}}(T))~ dk
\protect\label{EoS1cnT}
\end{eqnarray}

\begin{eqnarray}         
\hspace*{-2cm}P =-\frac{m^2}{8~C_{\sigma}}(1-Y^2)^2+\frac{m^2 B}{12~C_{\omega}~C_{\sigma}}(1-Y^2)^3
-\frac{C~ m^2}{16~ C_{\omega}^2 C_{\sigma}}(1-Y^2)^4 \nonumber \\
\hspace*{-2cm}+\frac{1}{2Y^2}C_{\omega} {\rho^2} + \frac{\gamma}{3\pi^2} \int_{0}^{\infty} \frac{k^4}{\sqrt{(k^2+{m^*}^2)}}(f(T)-\bar{f}(T))~ dk \nonumber \\+ \frac{\gamma}{6\pi^2} \sum_{\lambda= e,\mu^-} \int_{0}^{\infty} \frac{k^4}{ \sqrt{(k^2+{m_\lambda}^2)}}(f_{\lambda}(T)+\bar{f_{\lambda}}(T))~ dk
\protect\label{EoS2cnT} 
\end{eqnarray} 

 where, $f_{\lambda}(T)$ and $\bar{f_{\lambda}}(T)$ are the lepton and anti-lepton distribution functions.

  It is well-known that at low density near $\rho_0$, SNM is prone to undergo liquid-gas phase transition \cite{Bharat,Baldo1999,Wang2000,Malheiro,Nicotra,Sahu2004}. The transition temperature is called the critical temperature which is determined by the inflection point of the pressure and binding energy curves with respect to the total nucleon density viz,
  
\begin{eqnarray}
\frac{\partial P}{\partial \rho}\Bigg|_{T=T_c}=\frac{\partial^2 P}{\partial \rho^2}\Bigg|_{T=T_c}=0
\protect\label{Tc} 
\end{eqnarray}
 
 In case of $\beta$ stable matter at finite temperature, the last terms of eqs. \ref{EoS1cnT} and \ref{EoS2cnT} contribute due to the presence of leptons $\lambda= e,\mu^-$ with the individual lepton density as
 
\begin{eqnarray}
\rho_{\lambda}=\frac{\gamma}{2\pi^2} \int^{\infty}_0 dk ~k^2(f_{\lambda}(T)-\bar{f_{\lambda}}(T))  
\end{eqnarray}

where, $\lambda= e,\mu^-$ \nonumber

 Another important property in connection with matter at finite temperature is the free energy of the system. In terms of energy and entropy densities ($\varepsilon$ and $S$), the free energy density $F/\rho$ can be written as
\begin{eqnarray}
F/\rho=\varepsilon - TS
\protect\label{FAeq} 
\end{eqnarray}

 At $T=0$, the free energy density reduces to the energy density of the system.

 The entropy per volume or entropy density ($S/V$) can be obtained as 
\begin{eqnarray}
S/V=(P+\varepsilon-\sum_B \mu_B\rho_B)/T
\protect\label{SV} 
\end{eqnarray}

while and entropy per baryon($S$) is given by
\begin{eqnarray}
S=(P+\varepsilon-\sum_B \mu_B\rho_B)/\rho T
\protect\label{S} 
\end{eqnarray}

in the units of $k_B=1$.

 The speed of sound in nuclear medium is given by the first derivative of pressure with respect to energy density viz.
 
\begin{eqnarray}
{C_s}^2=\frac{dP}{d\varepsilon}
\protect\label{Cs} 
\end{eqnarray}

\subsection{The model parameter}

  In order to compute the EoS, one needs to specify the coupling constants. For SNM, there are four parameters $C_{\sigma},C_{\omega},B~ \&~ C$ while for ANM there is an additional parameter $C_{\rho}$. Their values are determined by reproducing the properties of SNM at saturation density $\rho_0$. The detailed procedure of obtaining these model parameters can be found in \cite{TKJ}.

 Bellow in table \ref{table-1}, I list the parameter set chosen for the present work along with the saturation properties. The parameter set is adopted from \cite{TKJ} (set 11 of \cite{TKJ}).

\begin{table}[ht!]
\begin{center}
\caption{Nuclear matter model parameters considered for the present work (adopted from \cite{TKJ}). The saturation properties and the scalar meson mass $m_{\sigma}$ yielded by this parameter set are also provided.}
\setlength{\tabcolsep}{15.0pt}
\begin{center}
\begin{tabular}{cccccccc}
\hline
\hline
\multicolumn{1}{c}{$C_{\sigma}$}&
\multicolumn{1}{c}{$C_{\omega}$} &
\multicolumn{1}{c}{$C_{\rho}$} &
\multicolumn{1}{c}{$B/m^2$} &
\multicolumn{1}{c}{$C/m^4$} &
\multicolumn{1}{c}{$m_{\sigma}$}\\
\multicolumn{1}{c}{($\rm{fm^2}$)} &
\multicolumn{1}{c}{($\rm{fm^2}$)} &
\multicolumn{1}{c}{($\rm{fm^2}$)} &
\multicolumn{1}{c}{($\rm{fm^2}$)} &
\multicolumn{1}{c}{($\rm{fm^2}$)} &
\multicolumn{1}{c}{(MeV)} \\
\hline
6.772  &1.995  & 5.285 &-4.274   &0.292  &510 \\
\hline
\hline
\multicolumn{1}{c}{$m^{\star}$/$m$}&
\multicolumn{1}{c}{$K$} & 
\multicolumn{1}{c}{$B/A$} &
\multicolumn{1}{c}{$J$} &
\multicolumn{1}{c}{$L_0$} &
\multicolumn{1}{c}{$\rho_0$} \\
\multicolumn{1}{c}{} &
\multicolumn{1}{c}{(MeV)} &
\multicolumn{1}{c}{(MeV)} &
\multicolumn{1}{c}{(MeV)} &
\multicolumn{1}{c}{(MeV)} &
\multicolumn{1}{c}{($\rm{fm^{-3}}$)} \\
\hline
0.85  &303  &-16.3   &32  &87  &0.153 \\
\hline
\hline
\end{tabular}
\end{center}
\protect\label{table-1}
\end{center}
\end{table}
 
 The values of saturation properties like symmetry energy coefficient ($J = 32$~MeV), saturation density ($\rho_0 = 0.153$~$\rm{fm^{-3}}$) and binding energy per particle ($B/A = -16.3$~MeV) for SNM, yielded by the model parameter show harmonious consistency with findings of \cite{j1,rmf1}. Compared to the results of \cite{l1}, the present parameter set gives a little larger value of the slope parameter ($L_0 = 87$~MeV). However, \cite{rmf1} suggests $L_0$ to be in the range (25 - 115) MeV. Moreover, recent co-relations between the symmetry energy and tidal deformability and radius of a 1.4 $M_{\odot}$ NS have suggested that $L_0$ can be $\sim$ 80 MeV \cite{Fattoyev,ZhenYuZhu} (comparable with that obtained with the present model). The nuclear incompressibility ($K = 303$~ MeV) yielded by the chosen parameter set, though consistent with findings of \cite{Stone2}, is larger than that reported in \cite{k1,k2,Garg} and that chosen recently in works like \cite{Burgio}. The parameter set for the present work is chosen from \cite{TKJ}. It can be seen from \cite{TKJ} that other than this parameter set chosen for this work, there are also other sets for the same model (reported in table 1 of \cite{TKJ}) that are obtained similarly by reproducing the SNM properties. Few such parameter sets (e.g. sets 13 and 14 of \cite{TKJ}) are also in accordance with the bounds on different SNM properties and the EoS obtained with them pass through heavy-ion collision data. In fact sets 13 and 14 of \cite{TKJ} also satisfy the bounds on incompressibility reported in \cite{k1,k2,Garg}. However, as discussed in \cite{TKJ,Sen2} that it can be seen from table 1 of \cite{TKJ} that higher ratio of the scalar to vector couplings ($C_{\sigma}/C_{\omega}$) and increasingly negative values of $B$ gives higher value of effective mass and lower nuclear incompressibility and consequently softer EoS \cite{TKJ} yielding low mass NS configurations that do not satisfy the maximum mass constraint of NSs at $T=0$. It is therefore appropriate that for the present work, I choose the parameter set with higher value of incompressibility ($K = 303$~ MeV) which gives comparatively stiffer EoS that satisfy the maximum mass constraint of NSs at $T=0$, similar to what was chosen in \cite{Sen,Sen2,Sen3,Sen4}.

 It is shown in \cite{TKJ,TKJ2,TKJ3} that the EoS for both SNM and PNM, obtained with the concerned parameter set, is also in accordance with the heavy-ion collision data \cite{hic}. However, for PNM the EoS passes through the soft band of heavy-ion collision data \cite{TKJ}. The reason is attributed to the high value of nucleon effective mass ($m^*=0.85~ m$) yielded by the present model compared to other RMF models \cite{rmf,rmf1} and the dominance of vector repulsive force at high densities, as discussed in \cite{Sahu2004,TKJ,Sen2}. In comparison with other RMF models like NL3 etc, the model presents soft EoS which softens more with the inclusion of hyperons and $\Delta$ baryons at high density \cite{Sen,Sen2,Sen3,Sen4}. Unlike various models \cite{Katayama,Sulaksono,Sahoo,Miyatsu,Miyatsu2} which produce stiff EoS and satisfy the 2 $M_{\odot}$ criteria of NSs mass \cite{Ant,Cromartie} even in the presence of hyperons and $\Delta$s, the present model do not \cite{Sen,Sen2,Sen3,Sen4} when such heavier baryons are included. Moreover, the appearance of these heavier baryons are much uncertain as they are largely controlled by the different couplings. Therefore, similar to \cite{Bharat,Sahu2004,Burgio2010}, the contributions of such exotic degrees of freedom to the properties of PNSs are not considered in the present work. 
 
 This model parameter (shown in table \ref{table-1}) has also been used to study a few nuclear matter properties at finite temperature at low density \cite{Sahu2004}. The same parameter set has also been used to describe the properties of hybrid stars in the presence of hyperons and $\Delta$ baryons (at $T=0$) in both static and rotational conditions \cite{Sen,Sen2,Sen3,Sen4}. Most of the model parameters being related to the VEV of the scalar field, the model provides very limited free parameters to adjust the SNM properties \cite{Sahu2004,TKJ}.

\subsection{Neutron Star Structure \& Properties}
\label{Structure}

 With the obtained EoS for the neutrino free $\beta$ equilibrated matter at different temperature, the structural properties of PNS are computed in static conditions using the Tolman-Oppenheimer-Volkoff (TOV) equations, given as \cite{tov,tov1}
 
\begin{eqnarray}
\frac{dP}{dr}=-\frac{G}{r}\frac{\left[\varepsilon+P\right ]
\left[M+4\pi r^3 P\right ]}{(r-2 GM)},
\label{tov1}
\end{eqnarray}

\begin{eqnarray}
\frac{dM}{dr}= 4\pi r^2 \varepsilon,
\label{tov2}
\end{eqnarray}

where, for a specified EoS and a given choice of central energy density $(\varepsilon_c)$, $M(r)$ denotes the gravitational mass enclosed within the star of radius $r~(=R)$. 

 The baryonic mass $M_B(r)$ \cite{Sen,Sen2,Sen4,Glendenning} and the surface redshift $Z_S$ \cite{Hong2016,Sen2,Glendenning} of PNS are respectively given as 
 
\begin{eqnarray} 
M_B(r)=\int_{0}^{R} 4\pi r^2 ~\varepsilon~ m_B \left(1 - \frac{2GM}{r}\right)^{1/2} dr
\label{barmass}
\end{eqnarray}

where, $m_B$ is the mass of baryon (nucleons in the present work) and
 
 
\begin{eqnarray}
Z_S = \Biggl(1 - \frac{2GM}{R}\Biggr)^{-1/2} - 1
\label{Z}
\end{eqnarray}

\section{Result and Discussions}
\label{Results}

\subsection{Symmetric nuclear matter at low density and finite temperature}

\begin{figure}[!ht]
\centering
\includegraphics[scale=0.7]{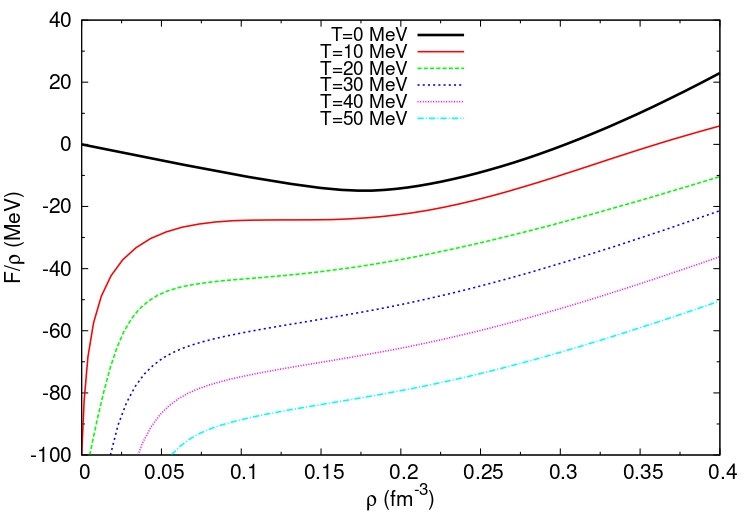}
\caption{\label{FA} Free energy per nucleon as a function of total baryon density for symmetric nuclear matter at different temperatures.}
\end{figure}

 Firstly, I consider SNM at low density and at finite temperature to compute the free energy per nucleon $F/\rho$ using equation \ref{FAeq}. Its variation with total nucleon density $\rho$ is shown in figure \ref{FA}.
 
\begin{figure}[!ht]
\centering
\includegraphics[scale=0.7]{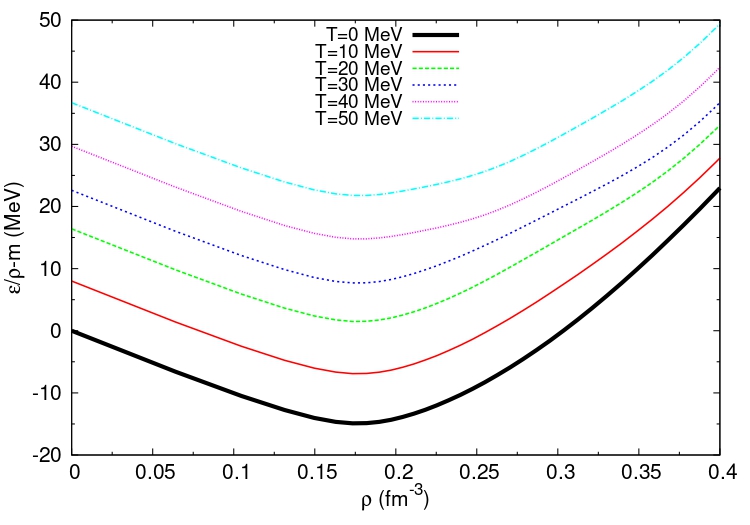}
\caption{\label{eb} Binding energy as a function of baryon density for symmetric nuclear matter at different temperatures.}
\end{figure}

 Since at $T=0$, free energy density reduces to the energy density of the system, therefore the curve for $T=0$ in figure \ref{FA} corresponds to the binding energy per nucleon curve obtained with SNM at $T=0$ displayed in figure \ref{eb}. The free energy density decreases with temperature. Consistent with the results of \cite{Tan,Nicotra,Lu,Burgio2010,Baldo1999,Wellenhofer2015}, I find that the decrease is more prominent at low baryon densities since the thermal effects on nuclear matter are more dominant at this density regime. With the increase of baryon density, the effect of temperature gradually reduces and therefore the free energy  isotherms almost converge at high density. I therefore choose to show the significant effect of temperature at the low density regime for both free energy density (figure \ref{FA}) and the binding energy per nucleon (figure \ref{eb}).
 
\begin{figure}[!ht]
\centering
\includegraphics[scale=0.7]{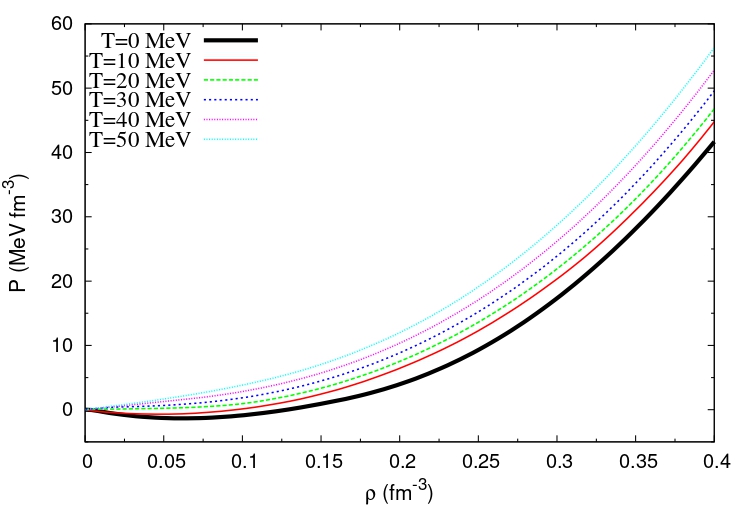}
\caption{\label{p} Pressure as a function of baryon density for symmetric nuclear matter at different temperatures.}
\end{figure}

 Next I calculate the binding energy per nucleon ($\varepsilon$) and pressure ($P$) of such a system. Their variations are studied with respect to total nucleon density $\rho$ and shown in figures \ref{eb} and \ref{p}, respectively. For better comparison, I have also shown the results at zero temperature ($T=0$).
 
 Both binding energy per nucleon and pressure decrease initially with nucleon density due to the dominance of the attractive scalar potential in this density domain. However, the curves of both the quantities rise after a certain density, depending on the temperature, due to the repulsive vector field potential that becomes dominant with increasing density. At $T=0$ the binding energy is -16.3 MeV for the chosen parameter set depicted in table \ref{table-1}. With increase in temperature, both $\varepsilon$ and $P$ increases and at a particular temperature (critical temperature $T_c$) they are no longer negative. This critical temperature is obtained using eq. \ref{Tc}. This indicates that the matter undergoes liquid-gas phase transition near the vicinity of saturation density. With increasing temperature, the minima of $\varepsilon$ not only tends to be positive but also shifts to a slightly higher density. For the present model considered, I obtain $T_c=16.54$ MeV, a temperature that starts separating the two phases. The obtained estimate of $T_c$ is consistent with the experimental bounds on the same ($T_c= 10 - 20$ MeV) \cite{Sharma,Kaijpper,Yang,Elliott,Takatsuka,Bharat,Sahu2004}. The corresponding values of critical density and critical pressure are obtained as $\rho_c=0.170$ fm$^{-3}$ and $P_c=4.31$ MeV~fm$^{-3}$. In table \ref{Tctable}, I compare the value of $T_c$ calculated in the present work with that obtained with a few other well-known theoretical models like NL3, G3, IU-FSU \cite{Bharat}, Walecka, ZM, ZM2 and ZM3 \cite{Malheiro}.
 
\begin{table}[ht!]
\caption{The critical temperature values $T_c$ for different theoretical models like NL3, G3, IU-FSU \cite{Bharat}, Walecka, ZM, ZM2 and ZM3 \cite{Malheiro}.}
\begin{center}
\begin{tabular}{ccccc}
\hline
\multicolumn{1}{c}{Models}&
\multicolumn{1}{c}{$T_c$ (MeV)}\\
\hline
This work   &16.54  \\
NL3         &14.60  \\
G3          &15.37  \\
IU-FSU      &14.50  \\
Walecka     &18.3   \\
ZM          &16.5   \\
ZM2         &15.5   \\
ZM3         &13.6   \\
\hline
\end{tabular}
\end{center}

\protect\label{Tctable}
\end{table}
 
\begin{figure}[!ht]
\centering
\includegraphics[scale=0.8]{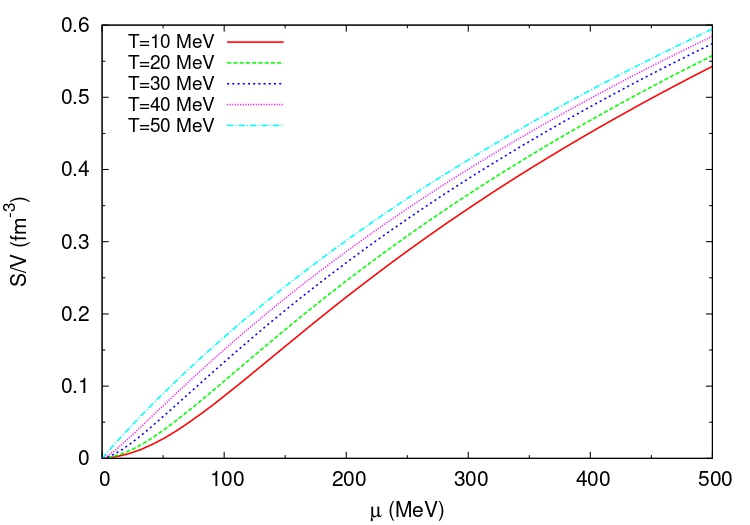}
\caption{\label{sv} Entropy density as function of chemical potential for symmetric nuclear matter at different temperatures.}
\end{figure}

\begin{figure}[!ht]
\centering
\includegraphics[scale=0.8]{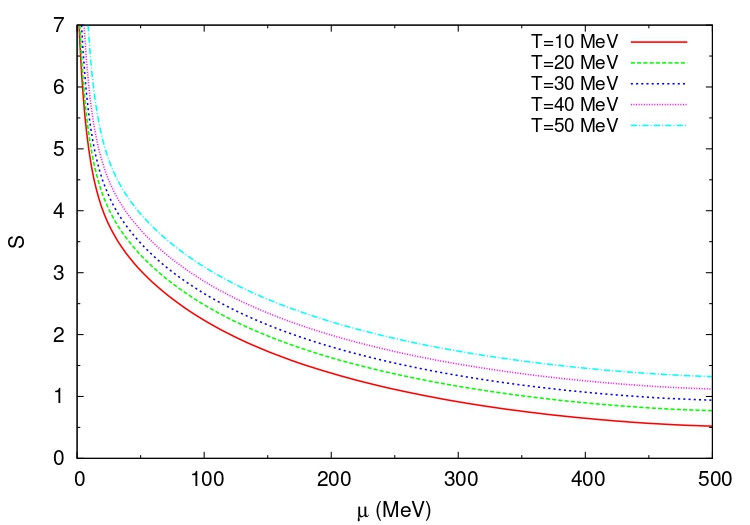}
\caption{\label{s} Entropy per baryon as a function of chemical potential for symmetric nuclear matter at different temperatures.}
\end{figure}

 The entropy density ($S/V$) and entropy per nucleon ($S$) are calculated next for SNM at different temperatures using eqs. \ref{SV} and \ref{S}, respectively. Their individual variation with respect to nucleon chemical potential at different temperatures are displayed in figures \ref{sv} and \ref{s}, respectively.

 I find that, consistent with the findings of \cite{Zhang2016,Tan,Burgio2010}, both entropy density and entropy per particle increases with the increasing value of temperature. However, at a given temperature, entropy density increases with nucleon chemical potential or total nucleon density while entropy per particle shows the opposite behavior. Clearly, at lower temperature $S$ decreases slowly as compared to that at higher temperature and at lower nucleon chemical potential $S$ decreases more sharply than at higher nucleon chemical potential. 

\subsection{Asymmetric nuclear matter at high density and finite temperature}

\begin{figure}[!ht]
\centering
\includegraphics[scale=0.8]{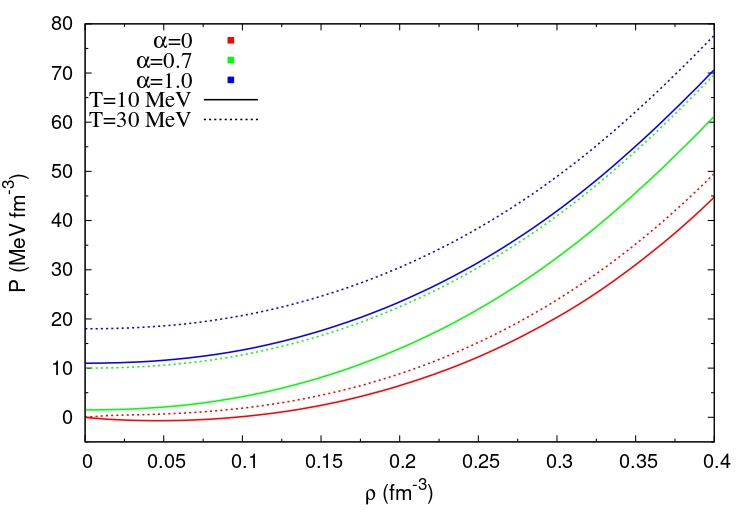}
\caption{\label{alpP} Variation of pressure with respect to density at $T=10~\&~ 30$ MeV for different values of $\alpha$. Red, green and blue curves correspond to $\alpha=$ 0, 0.7 and 1.0, respectively while solid and dashed curves correspond to $T=$ 10 and 30 MeV, respectively.}
\end{figure}

 As discussed in the formalism section, nuclear matter at high density becomes asymmetric i.e, the number of neutrons become much higher than that of protons. Therefore, I now treat matter at high density to be asymmetric and the contribution of the $\rho$ mesons now becomes crucial as they introduce the neutron-proton asymmetry in the system. Also the degree of asymmetry is determined by the asymmetric parameter $\alpha$, given by eq. \ref{alpha}. I fix {\textbf{$\alpha=0.7$}} to investigate the high density characteristics of ANM at finite temperature. In figure \ref{alpP}, I compare the variation of pressure with respect to density for $\alpha=$0 (SNM), 0.7 (ANM) and 1.0 (PNM) at low ($T=$ 10 MeV) and moderate ($T=$ 30 MeV) values of temperature.
 
  For a particular temperature, the increase in pressure is very rapid with the increase of the asymmetric parameter. At temperature as low as $T=10$ MeV, only SNM indicates phase transition and at moderately high temperature $T=30$ MeV, the pressure is positive for all the values of density and for all chosen values of $\alpha$. It is therefore that both temperature and the degree of asymmetry that play important role to determine the phase transition. Thus with the high value of $\alpha$ considered to explain neutron rich ANM, figure \ref{alpP} indicates no phase  transition for ANM. For SNM (red curves) the difference between the pressure at $T=10$ MeV and $T=30$ MeV is less compared to that obtained with $\alpha=0.7$ (green curves) or PNM (blue curves). This is because pressure increases more rapidly in the later two cases due to the combined effects of the $\rho$ mesons and the asymmetric parameter.
   
\begin{figure}[!ht]
\centering
\includegraphics[scale=0.8]{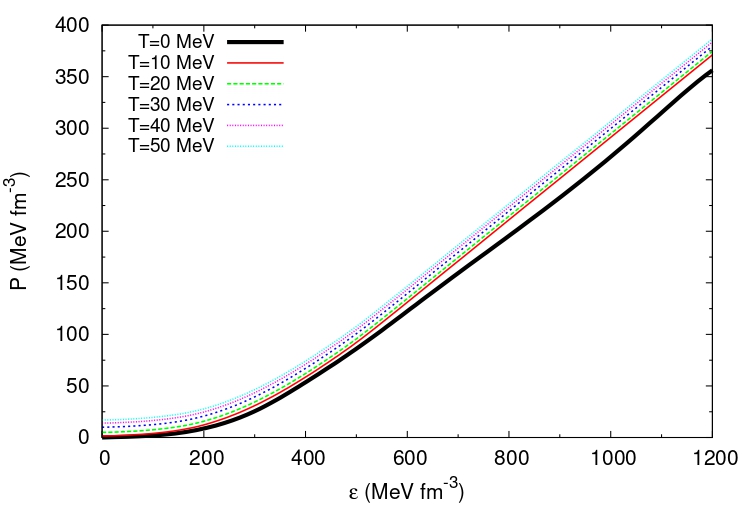}
\caption{\label{eos} Equation of State ($\varepsilon$ vs P) of asymmetric nuclear matter at different temperatures.}
\end{figure}

 In figure \ref{eos}, I present the equation of state i.e, variation of pressure with energy density for ANM at different values of temperature starting from $T=0$.

 There is no indication of liquid-gas phase transition with ANM as both energy density and pressure do not have negative values at any density, indicating that ANM is unbound for the chosen value of $\alpha$. It is seen that with increasing temperature, the EoS stiffens or the pressure increases for a given value of energy density as the anti-nucleons also contribute to the net thermal energy and pressure along with the nucleons \cite{Sahu2004}. Also the EoS is a function of nucleon momenta that rises with temperature.

\begin{figure}[!ht]
\centering
\includegraphics[scale=0.8]{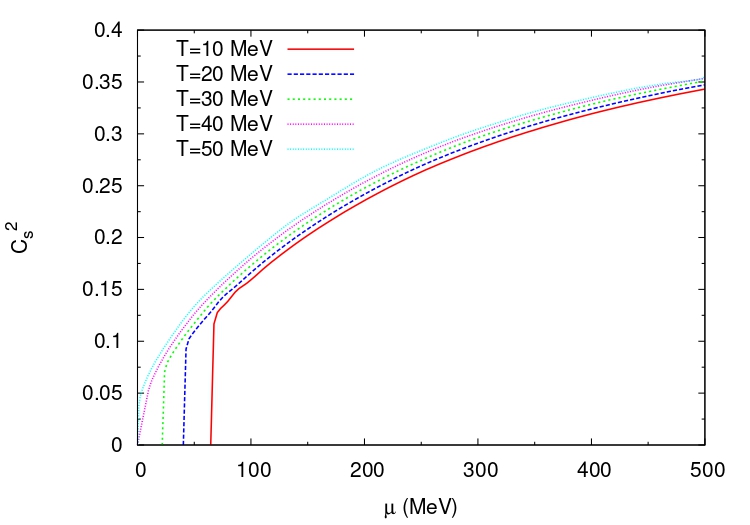}
\caption{\label{cs2} Speed of sound in nuclear medium as a function of chemical potential at different temperatures.}
\end{figure}

 The speed of sound in ANM is computed according to eq. \ref{Cs} and its variation with nucleon chemical potential for different temperatures is shown in figure \ref{cs2}.

 As expected, the speed of sound not only increases with nucleon chemical potential (total nucleon density) but also with temperature for a fixed value of nucleon chemical potential. It is interesting to note that at low temperature, the value of chemical potential for vanishing speed of sound is non-zero. It gradually decreases with the increase of temperature and finally becomes zero at $T=40$ MeV. At $T=50$ MeV it is noteworthy that the value of ${C_s}$ is non-zero at vanishing chemical potential. This is because the contribution to speed of sound is largely determined by both temperature and nucleon chemical potential. Thus at lower temperature the contribution to ${C_s}$ is dominantly from chemical potential which is therefore non-zero when ${C_s}=0$. As temperature increases one finds that the contribution to ${C_s}$ is gradually dominated by both temperature and chemical potential and thus the value of ${C_s}$ is non-zero at $T=50$ MeV for vanishing chemical potential.

\subsection{Static properties of proto-neutron stars}

 In order to calculate the structural properties of static PNS, I consider $\beta$ equilibrated matter with imposed the conditions of chemical equilibrium and charge neutrality, as described in the formalism section. I consider the neutrino untrapped scenario \cite{All,Chen,Bharat,Tan}.

 The EoS obtained for such matter at different temperature are individually subjected to the TOV eqs. \ref{tov1} and \ref{tov2} in order to calculate the structural properties of PNSs like gravitational mass $M$ ($M_{\odot}$) and radius $R$. The baryonic mass $M_B$ ($M_{\odot}$) and the surface redshift of the star are obtained from eqs. \ref{barmass} and \ref{Z}, respectively.
 
\begin{figure}[!ht]
\centering
\includegraphics[scale=1.0]{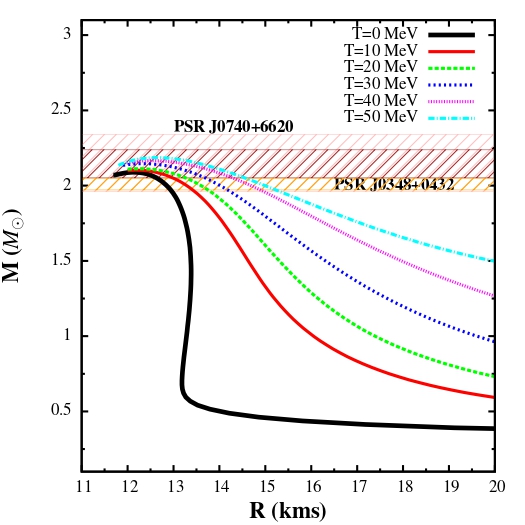}
\caption{\label{mr_all} Mass-radius relationship of proto-neutron stars in static conditions at different temperatures. Observational limits imposed from high mass pulsars like PSR J0348+0432 ($M = 2.01 \pm 0.04~ M_{\odot}$) \cite{Ant} (orange shaded region) and PSR J0740+6620 ($2.14^{+0.10}_{-0.09}~ M_{\odot}$ (68.3\% - brown shaded region) and $2.14^{+0.20}_{-0.18}~ M_{\odot}$ (95.4\% - pink shaded region)) \cite{Cromartie} are also indicated.}
\end{figure}

\begin{figure}[!ht]
\centering
\includegraphics[scale=0.8]{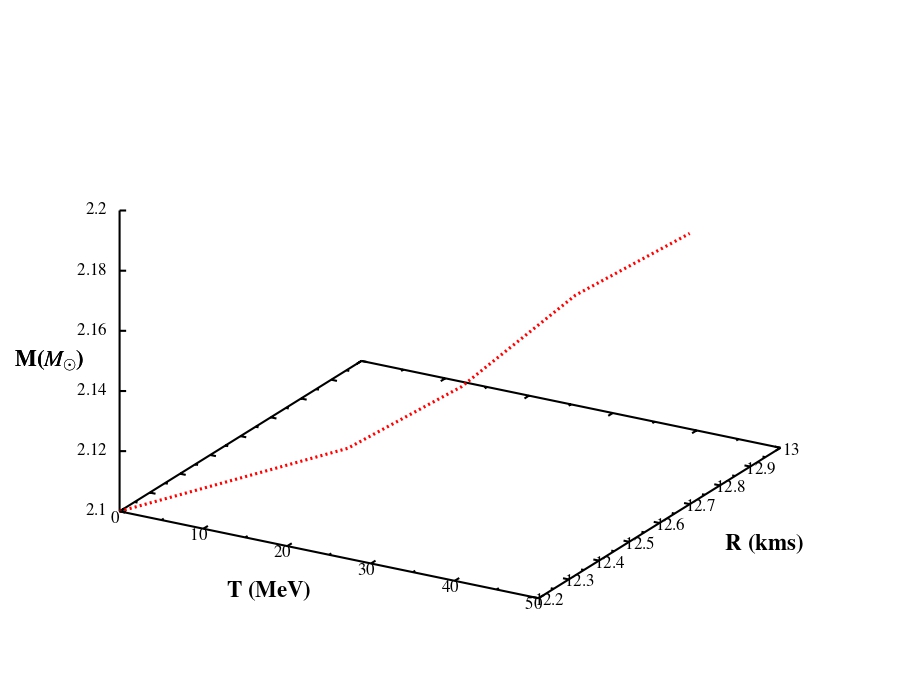}
\caption{\label{TMR} Variation of maximum mass and the corresponding radius of static proto-neutron stars with temperature.}
\end{figure}

 In figure \ref{mr_all}, I show the variation of gravitational mass with respect to radius of the star. To understand the explicit dependence of temperature on maximum mass and the corresponding radius of PNS, I show their variation in figure \ref{TMR}.

 I find that both maximum mass and the corresponding radius increase slightly with temperature since the EoS becomes stiffer with increasing temperature. Such a result is consistent with works like \cite{Yu,Bharat,Hong2016}. With EoS obtained at all temperatures, the maximum mass criteria of NSs from observational perspectives of massive pulsars like PSR J0348+0432 ($M = 2.01 \pm 0.04~ M_{\odot}$) \cite{Ant} and PSR J0740+6620 ($2.14^{+0.10}_{-0.09}~ M_{\odot}$ (68.3\%) and $2.14^{+0.20}_{-0.18}~ M_{\odot}$ (95.4\%)) \cite{Cromartie} are satisfied.

 In the present work the structural properties of PNS are calculated using EoS at constant temperature. However, \cite{Pons,BurgioNew} noted that with fast deleptonization, the density dependence of temperature in PNS no longer remains uniform. Under such circumstances, studying the entropy dependence of the structural properties of PNS becomes physically more justified \cite{BurgioNew,Burgio2011,Roark,Bharat,Dex,Tan,Burgio2010,Hong2016}. However, \cite{BurgioNew,Burgio2010} have already shown that in either cases, the maximum gravitational mass ($M$) is very less affected by the thermal effects. This result is quite consistent with that of the present work since I find a very slight increase in $M$ with respect to temperature. With the variation in temperature, the maximum gravitational mass varies from 2.10 $M_{\odot}$ to 2.19 $M_{\odot}$. Thus the increment is only in the second decimal place and becomes comparatively slightly more only when the temperature is as high as 40 or 50 MeV (as seen from table \ref{Stat.Prop}). This is because the nucleon momentum is dependent on temperature (section \ref{EOS_T}) and this in turn affects the pressure at finite temperature (eq. \ref{EoS2cnT}) and thereby the maximum mass of PNS. However, the effect being very feeble, it can be said that the PNS structural properties are controlled more by the composition and interactions considered than the thermal effects. The radius ($R$) corresponding to maximum mass also shows a very slight change with temperature (12.20-12.66) km. Therefore it appears that the structural properties of massive PNSs are quite weakly affected by thermal effects. However, with increasing values of temperature, low mass and moderately heavy PNSs are prone to have quite large values of radius compared to the cold NSs at $T=0$ (as seen from figure \ref{mr_all}), indicating that for such stars cooling plays a very important role to determine their size, volume and compactness.
 
 The baryonic mass ($M_B$) shows the same trend of dependence on temperature as gravitational mass lie in the range 2.41 $M_{\odot}$ to 2.64 $M_{\odot}$.
 
\begin{figure}[!ht]
\centering
\includegraphics[scale=1.0]{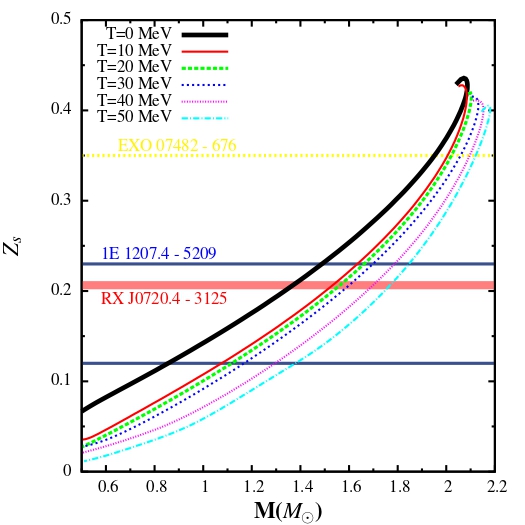}
\caption{\label{mZ_all} Surface redshift as a function of gravitational mass at different temperatures. Observational limits on redshift from EXO 07482-676 ($Z_s=0.35$) \cite{Cottam2002} (yellow dashed line), 1E 1207.4-5209 ($Z_s=(0.12-0.23)$) \cite{Sanwal} (area enclosed by indigo horizontal lines) and RX J0720.4-3125 ($Z_s=0.205^{+0.003}_{-0.006}$) \cite{Hambaryan} (red horizontal band) are also indicated.}
\end{figure}

 I also calculated the surface redshift ($Z_s$) of PNS for different temperatures according to eq. \ref{Z} and in figure \ref{mZ_all} I present the variation of $Z_s$ with respect to the gravitational mass.
 
 Consistent with the results of \cite{Hong2016}, I find that the maximum surface redshift decreases with increasing temperature (0.435 to 0.405). This is because the increase in temperature largely increases the radius of the star. It is clear from figure \ref{mZ_all} that the obtained estimates satisfy the observational constraints on redshift from different pulsars like EXO 07482-676 ($Z_s=0.35$) \cite{Cottam2002}, 1E 1207.4-5209 ($Z_s=(0.12-0.23)$) \cite{Sanwal} and RX J0720.4-3125 ($Z_s=0.205^{+0.003}_{-0.006}$) \cite{Hambaryan}.

 The various static properties of PNS at different temperatures are tabulated in table \ref{Stat.Prop}. 
 
\begin{table}[ht!]
\caption{The maximum gravitational mass $M$ ($M_{\odot}$), baryonic mass $M_B$ ($M_{\odot}$), radius $R$ (km) and the maximum surface redshift ($Z_s$) are displayed are displayed for corresponding temperature $T$ (MeV).}
\begin{center}
\begin{tabular}{ccccc}
\hline
\multicolumn{1}{c}{$T$}&
\multicolumn{1}{c}{$M$}&
\multicolumn{1}{c}{$M_{B}$} &
\multicolumn{1}{c}{$R$}&
\multicolumn{1}{c}{$Z_s$}\\
%
\multicolumn{1}{c}{(MeV)} &
\multicolumn{1}{c}{($M_{\odot}$)} &
\multicolumn{1}{c}{($M_{\odot}$)} &
\multicolumn{1}{c}{($km$)} &
\multicolumn{1}{c}{} \\
\hline
0   &2.10 &2.41 &12.20 &0.435 \\
10  &2.11 &2.49 &12.28 &0.427 \\
20  &2.12 &2.53 &12.40 &0.418 \\
30  &2.14 &2.55 &12.51 &0.414 \\
40  &2.17 &2.60 &12.58 &0.410 \\
50  &2.19 &2.64 &12.66 &0.405 \\
\hline
\end{tabular}
\end{center}

\protect\label{Stat.Prop}
\end{table}

\section{Conclusion}
\label{conclusion}

 I studied the finite temperature aspects of nuclear matter with the effective chiral model. For SNM I investigated the possibility of liquid-gas phase transition and the obtained estimate of critical temperature ($T_c=16.54$ MeV) in connection, is consistent with the range prescribed from experimental and empirical techniques. The free energy and entropy variation are also studied for SNM at different temperatures. I also examined the EoS and the variation of speed of sound in case of ANM. I found that both temperature and the nucleon chemical potential have dominant roles to play at different densities in order to determine the speed of sound in ANM.
 
 I also extended the work to determine the structural properties of PNSs like the gravitational and baryonic mass, radius and the surface redshift. I considered the neutrino free case and $\beta$ equilibrated matter as PNSM. The explicit dependence of these structural properties on temperature is examined by computing them at a wide range of temperature relevant to PNSs. Feeble increase in maximum gravitational mass and the corresponding radius is seen with respect to temperature while the radius of a low/moderately heavy PNS increases rapidly with temperature. I found that the estimates of maximum gravitational mass ($M=2.10-2.19~ M_{\odot}$) for all the chosen values of temperature are consistent with the observational bounds obtained from massive pulsars like PSR J0348+0432 and PSR J0740+6620. Also the predicted values of surface redshift from the present calculations ($Z_s=0.12-0.23$), are in par with that obtained from observational analysis of EXO 07482-676, 1E 1207.4-5209 and RX J0720.4-3125.

\ack
 
I am grateful to Dr. Sandeep Chatterjee, Department of Physical Sciences, Indian Institute of Science Education and Research Berhampur, for his useful suggestions and rigorous discussions regarding this work.


\end{document}